%% file: main.tex
\documentclass[a4paper,11pt]{article}
\usepackage{jinstpub} % for details on the use of the package, please see the JINST-author-manual
\usepackage{subfig}
\usepackage{orcidlink}
\title{\boldmath Comparative study and optimization of SDHCAL hadronic energy reconstruction methods}

\author[1]{T. Pasquier\note{Corresponding author.} \orcidlink{0009-0004-2267-8792}, }
\author{G. Grenier \orcidlink{0000-0002-1976-5877}, }
\author{ and I. Laktineh \orcidlink{0000-0002-8480-553X}}
\affiliation{IP2I Lyon, Université Lyon 1, Lyon, France}

\emailAdd{t.pasquier@ip2i.in2p3.fr}

\abstract{
We present a detailed study of hadronic shower energy reconstruction methods for the Semi-Digital Hadronic Calorimeter (SDHCAL) within the ILD detector concept, using the Particle Flow Algorithm (PFA) APRIL. Using samples of single $K^0_L$ and dijet ($u,d,s$) events, we compare linear, quadratic, split, and polynomial regression-based reconstruction formulas, focusing on their impact on linearity and resolution. The study also addresses angular corrections required in the barrel region due to non-perpendicular particle incidence. Results show that while all methods achieve good overall performance, the split method and the polynomial regression provide the best compromise across different energy regimes, offering improved resolution at low energies without compromising linearity at higher energies. For dijets, sensitivity to PFA confusion dominates the resolution at high energies. These findings highlight the potential of future improvements, notably the integration of precise timing information from the T-SDHCAL into APRIL, to further reduce confusion and enhance hadronic energy reconstruction for next-generation lepton colliders.
}


\begin{document}
\maketitle
\flushbottom

\input{section1}

\input{section2}
\input{section3}

\acknowledgments

The study is done within the framework of ILD Concept Group. The authors would like to thank the ILD Publication and Speakers Bureau and the ILD team at DESY for very useful comments and suggestions regarding this paper.

% This is the most common positions for acknowledgments. A macro is
% available to maintain the same layout and spelling of the heading.

% \paragraph{Note added.} This is also a good position for notes added
% after the paper has been written.

% % Bibliography

% %% [A] Recommended: using JHEP.bst file
\bibliographystyle{JHEP}
\bibliography{biblio.bib}

% %% or
% %% [B] Manual formatting (see below)
% %% (i) We suggest to always provide author, title and journal data or doi:
% %% in short all the informations that clearly identify a document.
% %% (ii) please avoid comments such as "For a review'', "For some examples",
% %% "and references therein" or move them in the text. In general, please leave only references in the bibliography and move all
% %% accessory text in footnotes.
% %% (iii) Also, please have only one work for each \bibitem.

% \begin{thebibliography}{99}

% \bibitem{a}
% Author,
% \emph{Title},
% \emph{J. Abbrev.} {\bf vol} (year) pg.

% \bibitem{b}
% Author,
% \emph{Title},
% arxiv:1234.5678.

% \bibitem{c}
% Author,
% \emph{Title},
% Publisher (year).

% \end{thebibliography}
\end{document}

%% file: section1.tex
\section{Introduction}

Future high-energy colliders, also referred to as Higgs factories, will require high-precision measurements to test the robustness of the Standard Model and search for signs of new physics~\cite{ILC_TDR,CEPC_TDR,FCC_CDR}. To meet this challenge, various Particle Flow Algorithms (PFA)~\cite{Brient:2001fow} - such as PandoraPFA~\cite{Thomson_2009}, ARBOR~\cite{ruan2014arbornewapproachparticle} and APRIL~\cite{Li_2020} - have been developed. The PFA approach consists in reconstructing and identifying each individual particle in an event by combining tracking and calorimetric information, assigning charged particle momenta to the tracker measurement and using calorimetric information for neutral particles. The use of such algorithms is enabled by high-granularity calorimeters that allow for the efficient separation of energy depositions of close-by particles. 

The Semi-Digital Hadronic CALorimeter (SDHCAL)~\cite{CommissioningProtoSDHCAL, FirstResultsSDHCAL}, developed within the CALICE collaboration~\cite{CALICE:2012ami}, is one of the proposed high-granularity HCAL for future detectors such as ILD~\cite{ILD_detector, ILDConceptGroup:2020sfq}. 

This paper focuses on the hadronic energy reconstruction of the SDHCAL within the ILD detector concept, specifically using the \texttt{ILD\_l2\_v02} configuration from the \texttt{k4geo} package~\cite{k4geo_ref}, which implements the SDHCAL in the Videau geometry (see section~\ref{simulation_section}).\footnote{Available at \url{https://github.com/key4hep/k4geo/tree/main/ILD/compact/ILD_l2_v02}.}

The study first addresses the need for angular corrections in the energy reconstruction process and the SDHCAL calibration procedure. It then presents a comparison of several reconstruction formulas to assess their impact on hadronic energy reconstruction using both $K^{0}_{L}$ and dijet events.

\subsection{The SDHCAL}
\label{SDHCALIntro}

\begin{figure}[h]
    \centering
    \includegraphics[width=0.7\linewidth]{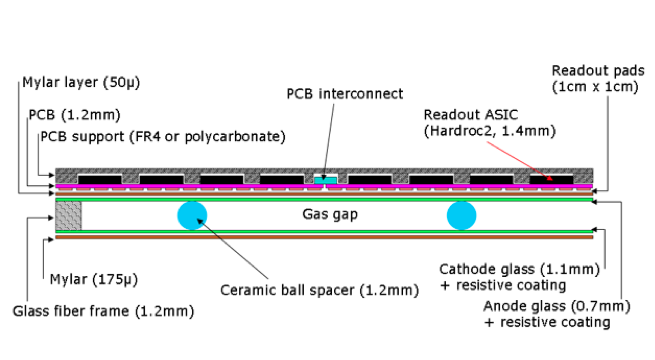}
    \caption{Schematic cross-section view of a SDHCAL GRPC.}
    \label{sdhcal_grpc_fig}
\end{figure}

The SDHCAL is a high-granularity sampling hadronic calorimeter developed by the CALICE collaboration. A $1\,\text{m}^3$ technological prototype  has been designed, constructed, and tested several times in test beam at CERN to validate its performances~\cite{CommissioningProtoSDHCAL, FirstResultsSDHCAL}. It consists of 48 layers combining stainless steel absorber plates and Glass Resistive Plate Chambers (GRPC) with embedded readout electronics (see figure~\ref{sdhcal_grpc_fig}). Each GRPC is coupled to an Active Sensor Unit (ASU) segmented into $96\times96$ copper pads of $1~\text{cm}^2$ on one side, ensuring high granularity, and hosting HARDROC2 ASICs~\cite{HARDROC2_reference} on the other. Each HARDROC2 chip is connected to 64 pads in an $8\times8$ array, making a total of 144 ASICs per layer. Signals induced by particles ionizing the 1.2~mm gas gap are read out as 2-bit information, providing three thresholds in a range from 100~fC to 15~pC. 

The mechanical structure integrates both the absorber and the support, minimizing dead zones and making the prototype well-suited for PFA applications. Each layer is enclosed in a thin cassette made of two stainless steel plates that protect the GRPC and the ASU and also contribute to the total absorber thickness of the calorimeter. The 48 cassettes are inserted into this self-supporting structure, allowing for a compact and easily assembled design.

A new version of this prototype, the T-SDHCAL~\cite{TSDHCAL}, is currently under development with the aim of replacing GRPCs by Multigap Resistive Plate Chambers (MRPCs), providing an excellent time resolution to improve the reconstruction and separation of the showers within the PFA framework. 

\subsection{The International Large Detector}

The International Large Detector (ILD)~\cite{ILD_detector, ILDConceptGroup:2020sfq} is one of the detector concepts proposed for next-generation colliders operating as Higgs factories. It has been designed to fully exploit the PFA approach by combining high-granularity calorimetry with precise tracking. The ILD detector concept features a layered structure. It first consists of a silicon vertex detector located near the interaction point, and a Time Projection Chamber (TPC) able to achieve a point resolution better than $100\,\mu\mathrm{m}$ for the complete drift and a double hit resolution
less than 2\,mm under a 3.5~T magnetic field. The TPC is complemented by silicon tracking layers forming the so-called Silicon Envelope: two barrel components, the Silicon Inner Tracker (SIT) and the Silicon External 
Tracker (SET), and the Forward Tracking Detector (FTD). These are followed by a finely segmented electromagnetic calorimeter (ECAL), and a highly granular hadronic calorimeter (HCAL). All these subdetectors are enclosed within a solenoid and surrounded by an instrumented iron return yoke acting as a muon detector. The ILD detector concept is defined in two size configurations, commonly referred to as the large and small models. These versions share the same overall design philosophy but differ mainly in their dimensions and magnetic field strength, with the large model operating at 3.5~T and the small one at 4~T. A magnetic field of 2~T is also foreseen in the context of FCC-ee operation.

The design aims to achieve a jet energy resolution of $\sigma_E/E \sim 3-4\%$ 
for jets of 100~GeV~\cite{ILD_detector, ILDConceptGroup:2020sfq}, enabling the separation of 
hadronically decaying W and Z bosons, as well as efficient particle identification and precise momentum measurement that will be key ingredients for Higgs and electroweak precision studies at future electron–positron colliders. Different technological options are being investigated for each subsystem, reflecting the modularity and flexibility of the ILD detector concept. In particular, the SDHCAL is one of the proposed HCAL technologies.

%% file: section2.tex
\section{SDHCAL calibration and corrections}

\subsection{Simulation and reconstruction}
\label{simulation_section}

The ILD detector geometry is described using DD4hep~\cite{dd4hep_citation} and the \texttt{k4geo} package~\cite{k4geo_ref}, which provide the \texttt{ILD\_l2\_v02} configuration used in this study. The simulation is performed with \texttt{ddsim}, the DD4hep-based simulation driver, which interfaces with GEANT4~\cite{GEANT4} for the physics processes. All simulated events in this study are generated using the \texttt{QGSP\_BERT} physics list. Additional tests were carried out with the \texttt{FTFP\_BERT} physics list and yielded similar results. Digitization and reconstruction are handled within the Marlin framework~\cite{Marlin_citation}, as part of the iLCSoft v02-03-03 software ecosystem.\footnote{Available at \url{https://github.com/iLCSoft/iLCInstall/releases/tag/v02-03-03}.} The results presented in this paper are obtained using simulations of the \texttt{ILD\_l2\_v02} detector model. This model is one of the proposed configurations of the ILD detector concept and includes the SDHCAL as the hadronic calorimeter. The detailed geometry of the SDHCAL cassette is fully implemented in the detector model. In this configuration, the SDHCAL consists of 48 layers and is located outside the inner tracking system and the SiW ECAL~\cite{SiW_ECAL}. As the \texttt{ILD\_l2\_v02} detector model corresponds to one of the large ILD configurations, the barrel HCAL has an inner radius of 2058~mm, an outer radius of 3345~mm and a total length of 4700~mm along the z-axis~\cite{ILDConceptGroup:2020sfq}.  

\begin{figure}[h]
    \centering
    \includegraphics[width=0.44\linewidth]{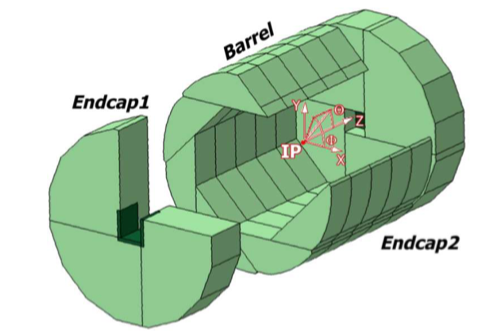}
    \caption{3D-view of the SDHCAL geometry for the ILD\_l2\_v02 configuration.}
    \label{videau_full}
\end{figure}

\begin{figure}[h]
    \centering
    \includegraphics[width=0.55\linewidth]{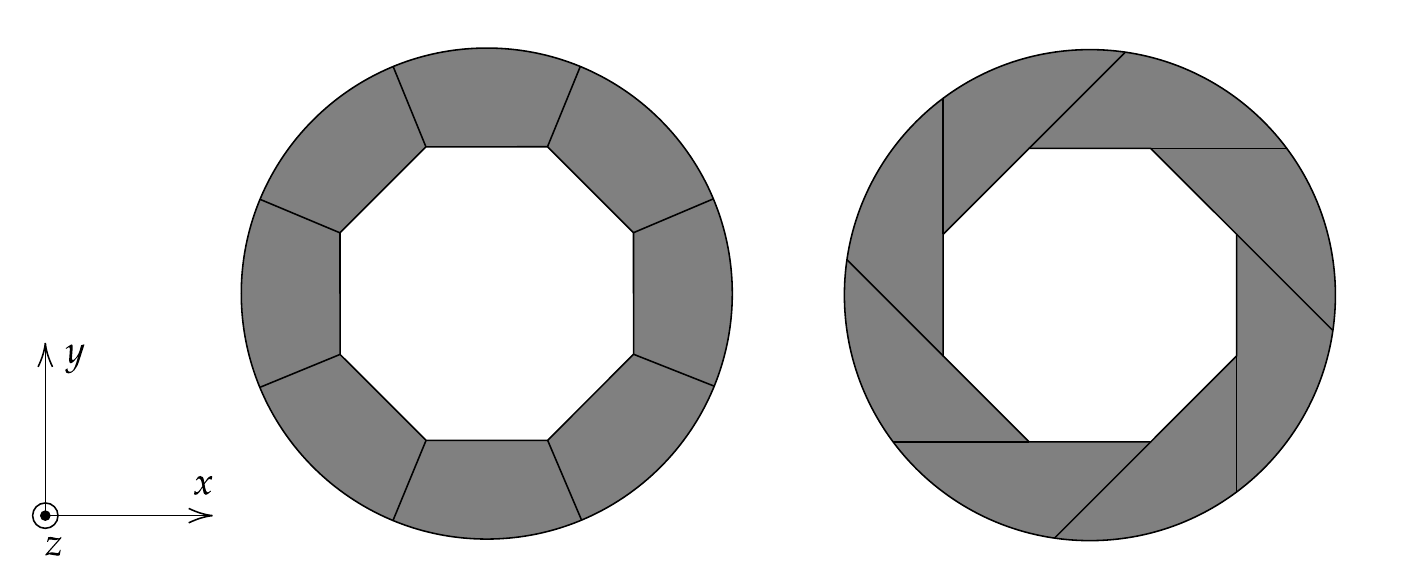}
    \caption{xy-view of the SDHCAL TESLA geometry (left) and Videau geometry (right).}
    \label{Tesla_vs_Videau}
\end{figure}

The barrel region of the HCAL is made of five wheels aligned along the z-axis, each segmented into eight modules, as shown in figure~\ref{videau_full}. The particular shape of the modules is called the Videau geometry. This geometry minimizes uninstrumented regions encountered by particles originating from the interaction point. In the TESLA geometry (another possible HCAL design), gaps between modules are aligned with straight lines pointing to the interaction point, so a particle emitted along such a direction could pass entirely through a gap and be lost. In the Videau geometry, the module boundaries are overlapping with respect to the interaction point, significantly reducing this effect, as illustrated in figure~\ref{Tesla_vs_Videau}.

\begin{figure}
    \centering
    \includegraphics[width=0.6\linewidth]{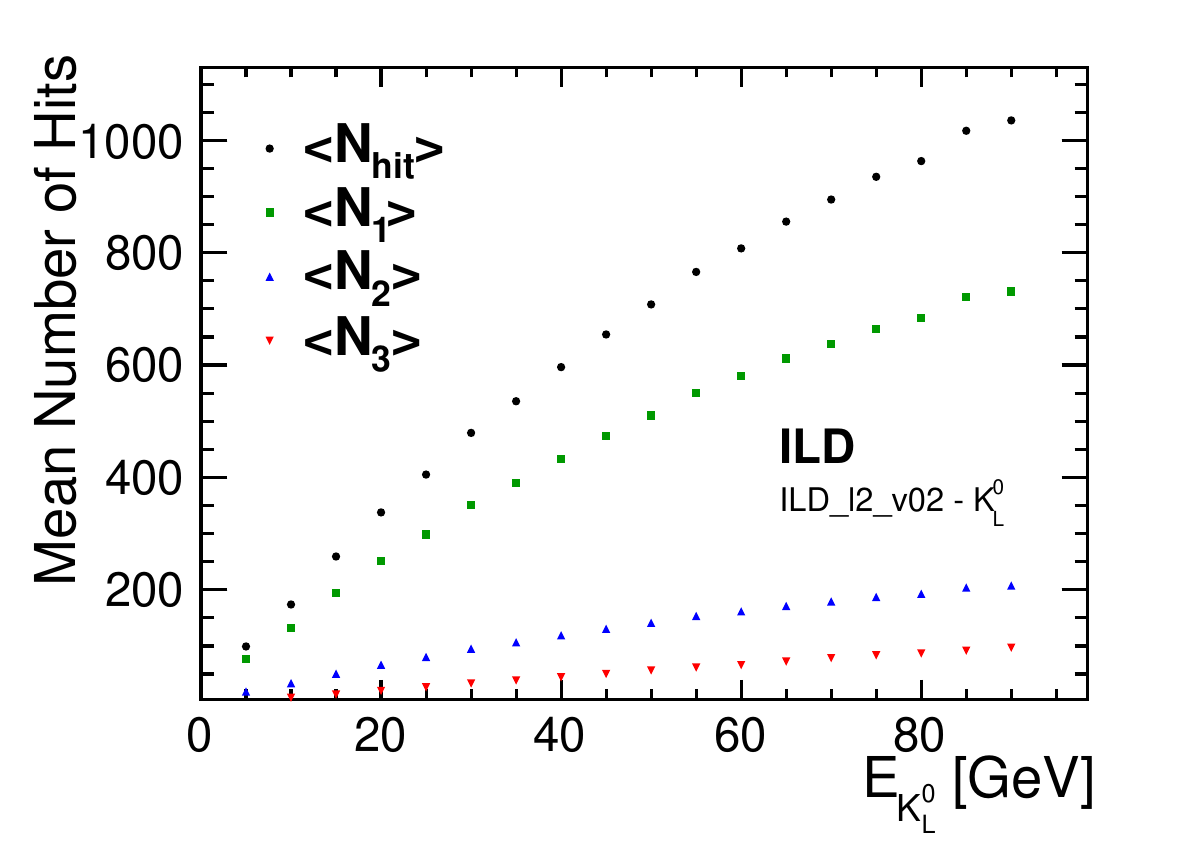}
    \caption{Average number of digitized hits in the SDHCAL corresponding to the first threshold (green squares), the second threshold (blue triangles), the third threshold (red triangles), and the total number of hits (black circles) as a function of the $K^{0}_L$ energy in the ILD\_l2\_v02 simulation.}
    \label{N1N2N3vsEnergy}
\end{figure}

As explained in section~\ref{SDHCALIntro}, the SDHCAL provides a semi-digital response with three thresholds. In the simulation, this response is reproduced through a digitization process that converts the charge deposited by particles of the shower into signals corresponding to these three thresholds, producing so-called hits~\cite{Deng_2016}. The thresholds used in the simulation are fixed to 114~fC, 6.12~pC and 16.83~pC to best reproduce the multiplicity and efficiency obtained with the prototype in test beam~\cite{garillot:tel-02141420}. The three thresholds provide information on the local particle density within a shower. A minimum ionizing particle (MIP) typically induces an average charge of about 1.2~pC with a rather large spread~\cite{Ran_2014}. As a result, isolated particles generally exceed only the first threshold, while regions with several overlapping charged particles are more likely to exceed the second or third thresholds. When a particle triggers a shower in the detector, it is possible to determine $N_1$, $N_2$, and $N_3$, where $N_1$ corresponds to the number of hits above the first threshold and below the second, $N_2$ to hits above the second threshold and below the third, and $N_3$ to hits above the third threshold. The average values of $N_1$, $N_2$, $N_3$ and of the total number of hits $N_{\text{hit}}$ ($N_{\text{hit}}=N_1+N_2+N_3$) in the SDHCAL for samples of single $K^{0}_L$ in the ILD\_l2\_v02 detector model are shown in figure~\ref{N1N2N3vsEnergy}. 

Reconstruction is done by the PFA by grouping calorimeter hits in clusters. The reconstructed energy is then expressed as the sum of the ECAL and HCAL contribution. For hits in the ECAL, the energy contribution is the sum of hit energies. For those in the SDHCAL, the energy contribution is computed as a function of $N_1$, $N_2$ and $N_3$, as detailed in section~\ref{Calibration_procedure}.

\subsection{Angular corrections}
\label{angular_correction_chapter}

A preliminary study was first performed using the standard PandoraPFA calibration procedure for ILD,\footnote{Available at \url{https://github.com/PandoraPFA/LCPandoraAnalysis}.} which relies on samples of single long-lived neutral kaons ($K^{0}_L$), with an energy of 20~GeV and generated with a uniform angular distribution, that provide a clean probe of the hadronic calorimetric response, without tracking or magnetic field effects. Using this procedure, the reconstructed energy was studied separately in the endcap and in the barrel regions of the \texttt{ILD\_l2\_v02} detector model. 

The particle energy was well reconstructed in the endcap, where particles travel almost perpendicularly to the layer surfaces, with $E_\mathrm{reco,endcap}=20.36\pm0.08~\mathrm{GeV}$, while a significant underestimation was observed in the barrel where the incidence angle of the particle can be larger, with $E_\mathrm{reco,barrel}=18.89\pm0.06~\mathrm{GeV}$. 

These results indicate that the SDHCAL is correctly calibrated with the parameters listed in table~\ref{Tab_linear}, described in section~\ref{Calibration_procedure}, but requires an additional correction in the barrel. 

\begin{figure}[htbp]
  \centering
  % Figure a
  \subfloat[]{
    \includegraphics[width=0.48\textwidth]{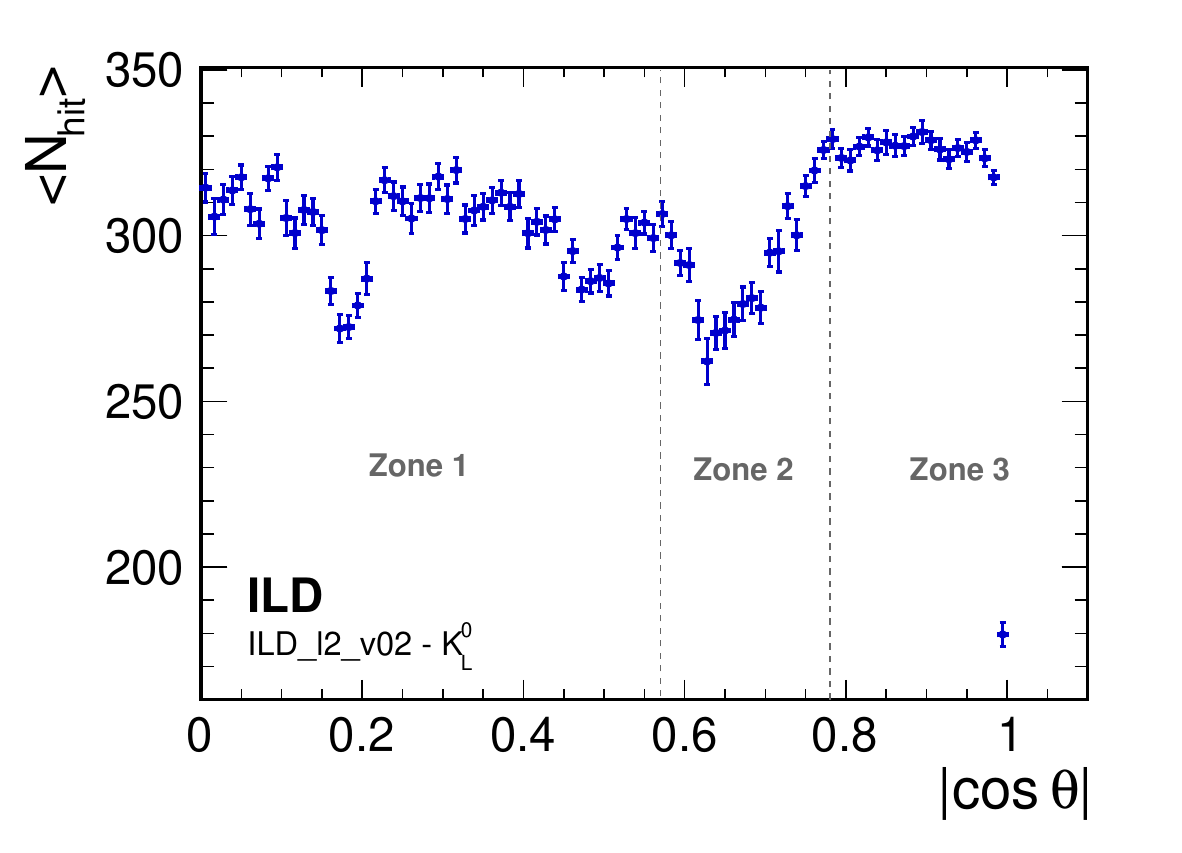}
    \label{SubfigCosTheta}
  }
  \hfill
  % Figure b
  \subfloat[]{
    \includegraphics[width=0.48\textwidth]{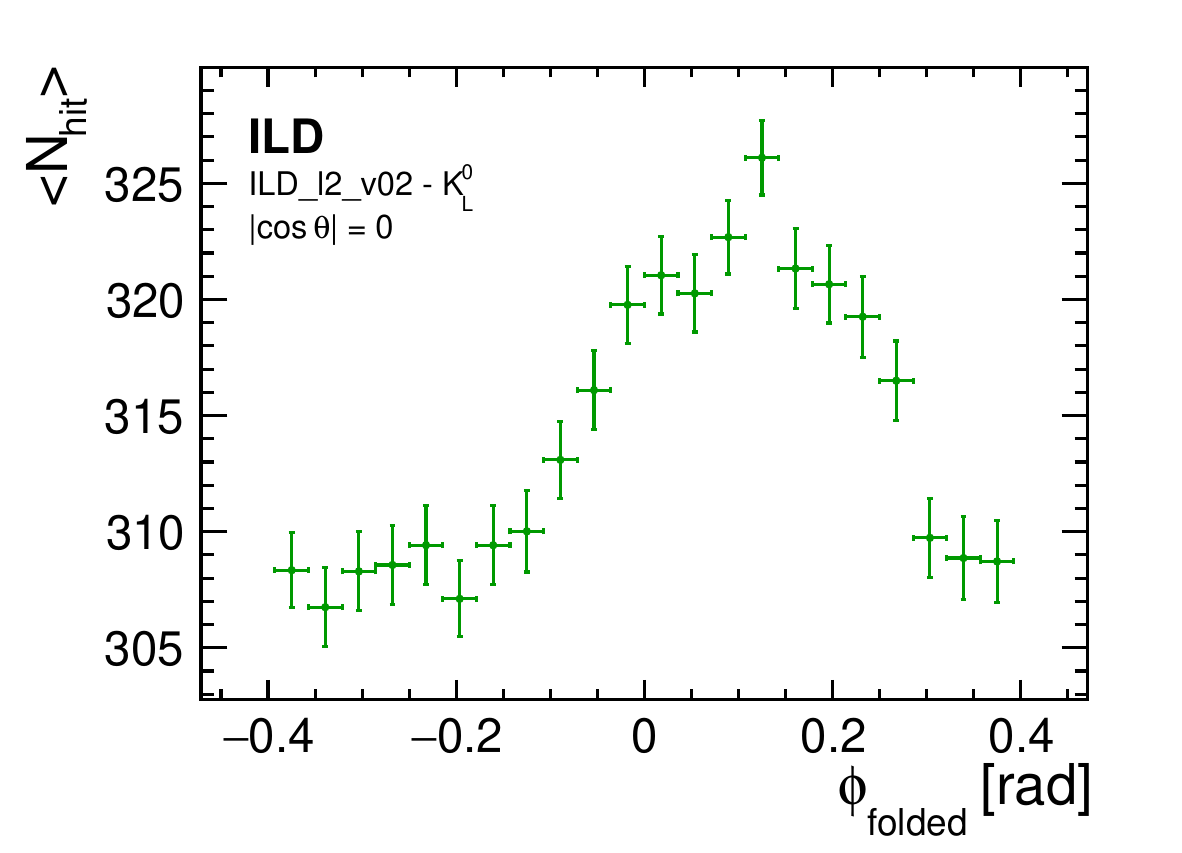}
    \label{SubfigPhi}
  }
  \caption{Mean number of hits in the SDHCAL as a function of $|\cos \theta|$ (left) and as a function of $\phi_{\mathrm{folded}}$ (right) for 20~GeV single $K^0_L$.}
  \label{fig_correction_hits}
\end{figure}

To investigate the origin of this underestimation, the mean number of hits recorded in the SDHCAL was studied as a function of $|\cos\theta|$, where $\theta$ is the angle between the $K^0_L$ momentum and the z-axis, and of the folded azimuthal angle $\phi_{\mathrm{folded}}$, for 20~GeV single $K^0_L$ events, as shown in figure~\ref{fig_correction_hits}. For figure~\ref{fig_correction_hits}\subref{SubfigCosTheta}, Zone 1 corresponds to the barrel only, Zone 2 to the transition between barrel and endcap, and Zone 3 to the endcap only. For figure~\ref{fig_correction_hits}\subref{SubfigPhi}, a dedicated sample of single $K^{0}_L$ events with an energy of 20~GeV has been generated with $|\cos\theta|=0$ and a uniform $\phi$ distribution. The azimuthal angle $\phi$ is computed from the projection of the $K^0_L$ momentum onto the transverse plane. In order to exploit the rotational symmetry of the calorimeter, $\phi$ is folded modulo $\frac{\pi}{4}$, corresponding to the angular periodicity of the detector geometry. The folded angle is therefore restricted to the interval $[-\pi/8, +\pi/8]$.

For $|\cos\theta|\simeq0$, figure~\ref{fig_correction_hits}\subref{SubfigCosTheta} shows that there are on average 315 hits in the SDHCAL. In the endcap (Zone 3), the mean number of hits is 325. This difference originates from the dependence on the azimuthal angle $\phi$, which only affects the barrel geometry. Due to the particular geometry of the HCAL modules in the Videau geometry, a particle traveling perpendicularly to the first module layers may penetrate deeply enough to reach the adjacent module which is tilted by an angle of 45°. In figure~\ref{fig_correction_hits}\subref{SubfigPhi}, when the $\phi$ angle is small, the number of hits in the barrel is around 320--325, which is coherent with the value found in the endcap. As the $\phi$ angle increases, this number of hits drops to 305--310, explaining why the mean number of hits in the barrel is below the one in the endcap.

This behavior is directly related to the sampling fraction of the SDHCAL. At larger incidence angles, fewer sensitive layers are crossed per unit path length, resulting in a reduced sampling of the shower by the active layers. Since the reconstructed hadronic energy is proportional to the number of SDHCAL recorded hits, this reduction in sampling directly translates into a systematic underestimation of the reconstructed energy in the barrel.
The same behavior occurs with $\cos\theta$ but is hidden in the several drops in the number of hits that can be observed in figure~\ref{fig_correction_hits}\subref{SubfigCosTheta} around $|\cos\theta|\simeq0.2$ and $|\cos\theta|\simeq0.5$. This effect is due to the non instrumented gaps between the barrel wheels and requires a specific correction which is not discussed here.

\begin{figure}[h]
    \centering
    \includegraphics[width=0.65\linewidth]{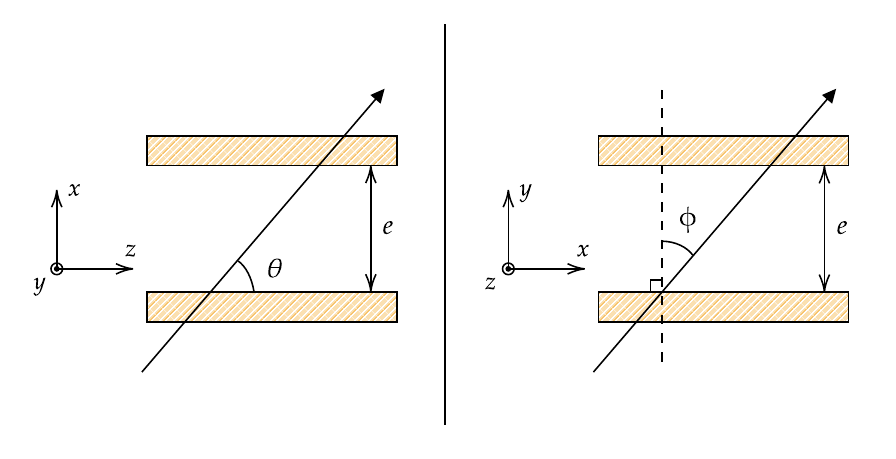}
    \caption{Schematic illustration of the geometrical effect associated with the angular correction applied in the SDHCAL barrel. Due to the particle incidence angle, the path length through a layer of thickness $e$ becomes $e/\sin\theta$ or $e/\cos\phi$, depending on the considered direction.}
    \label{schema_angular_correction}
\end{figure}

To account for the path length variation due to the particle's incidence angle, a purely geometrical correction is implemented by applying angular weights to the number of hits entering the reconstruction formula, with correction factors depending on the cluster's $\theta$ and $\phi$ angles. For a given cluster, $\theta$ is defined as the angle between the direction from the interaction point to the cluster centroid and the detector’s z-axis (beam axis). On the other hand, $\phi$ is defined as the angle between the normal vector of the calorimeter layers and the projection of the cluster direction onto the xy-plane. It should be noted that this definition differs from the global azimuthal angle used in figure~\ref{fig_correction_hits}\subref{SubfigPhi}. In the correction procedure, $\phi$ corresponds to a local incidence angle defined at the module level using reconstructed cluster information only. Since the shower may extend over several modules with different orientations, the $\phi$ angle varies from one module to another. In this case, the correction is computed separately for each module and the final corrected number of hits, $N_{i_{\text{corr}}}$, corresponds to the sum of contributions of all the concerned modules. 

\begin{figure}[htbp]
  \centering
  % Figure a
  \subfloat[Before angular corrections\label{BeforeCorrection}]{
    \includegraphics[width=0.48\textwidth]{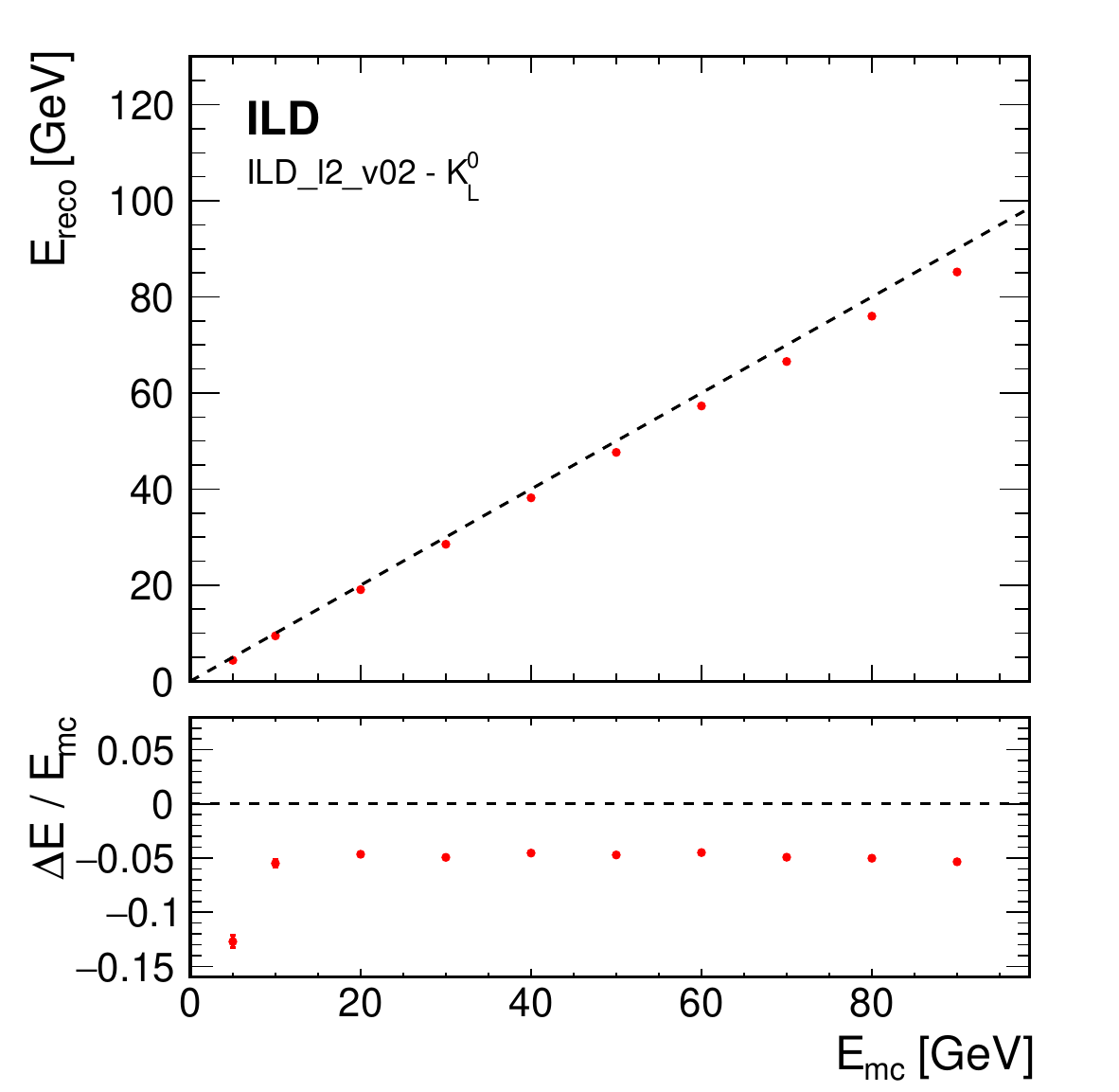}
  }
  \hfill
  % Figure b
  \subfloat[After angular corrections\label{AfterCorrection}]{
    \includegraphics[width=0.48\textwidth]{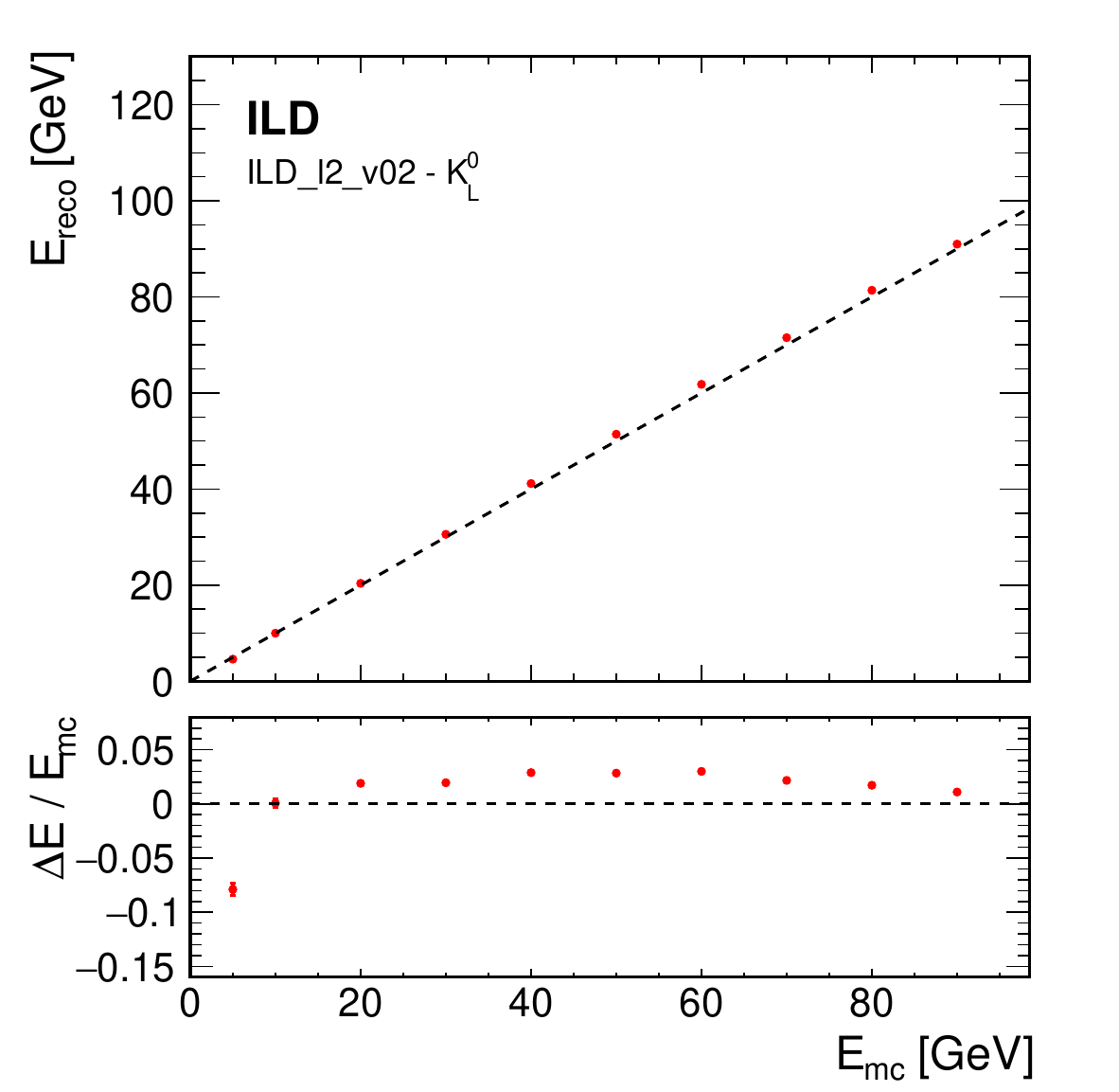}
  }
  \caption{Reconstructed energy and linearity for single $K^0_L$ particles sent with various angles into the barrel with $|\cos \theta| < 0.7$ and reconstructed with APRIL. Error bars are smaller than the marker size.}
  \label{fig_correction_angulaire}
\end{figure}

The geometrical effect underlying the correction factors is illustrated in figure~\ref{schema_angular_correction}.
The corrected number of hits for each threshold is finally given by:

\begin{equation}
\label{corrected_formula}
  N_{i_{\text{corr}}} = N_{i} + N_{i_{\text{barrel}}}\left( \frac{1}{\sin\theta} - 1 \right) + \sum_{\text{module}}N_{i_{\text{barrel, module}}}\left( \frac{1}{\cos(\phi_{\mathrm{module}})} - 1 \right) \text{ with } i=1,2,3.
\end{equation}

To further quantify the effect of the angular correction, a dedicated study was performed using single $K^{0}_L$ samples generated with a uniform angular distribution in the barrel with $|\cos \theta| < 0.7$, with energies going from 5~GeV to 90~GeV, and reconstructed using the default ILD reconstruction configuration and the quadratic formula described in section~\ref{Quad_reco}. For each energy point, 10\,000 events were simulated.

Before applying any angular correction, the reconstructed energy is systematically underestimated with respect to the true energy, as shown in figure~\ref{fig_correction_angulaire}\subref{BeforeCorrection}, confirming the need for a correction.

When applied, the correction reduces the systematic underestimation of the reconstructed energy observed in the barrel and mitigates the dependence on the incidence angle, as illustrated in figure~\ref{fig_correction_angulaire}\subref{AfterCorrection}. For energies above 20~GeV, a slight overestimation at the level of 2--3\% is observed. This effect is interpreted as a residual bias of the overall reconstruction and correction procedure. 
At 5~GeV, the reconstructed energy remains slightly underestimated, although moderately improved by the angular correction. This behavior is currently under investigation.

\subsection{Calibration dataset}
\label{Calibration_dataset}

The calibration procedure of the formula's parameters is performed on samples of $K^{0}_L$ sent individually in the barrel from the detector center with an initial momentum aligned with the x-axis (see figure~\ref{Tesla_vs_Videau}). In this configuration, particles enter the calorimeters with a direction normal to the detector layers of the first HCAL module. 10,000 events are simulated for each energy between 5~GeV and 90~GeV in steps of 1~GeV.  
In order to calibrate the intrinsic response of the HCAL, only events depositing energy exclusively in the HCAL are retained, rejecting those with hits in the ECAL. A containment requirement is then applied to reject events with activity in the outermost layers of the HCAL, based on a radial criterion consistent with the detector geometry, to ensure that the full energy of the $K^{0}_{L}$ is measured. After these selections, the final calibration dataset contains 71437 events. Although this represents a fraction of the initially simulated sample, it remains a sufficiently large dataset. The selection is intentionally designed to isolate well-contained hadronic showers and minimize the impact of leakage and upstream interactions. The robustness of the calibration is subsequently assessed on more realistic event topologies, including dijet events, as discussed in section~\ref{Results}.

Due to the particular geometry of the ILD\_l2\_v02 configuration, particles sent straight into the barrel may travel deep enough in the HCAL to deposit energy in the second module, which is tilted by 45° with respect to the first one, as shown in figure~\ref{Tesla_vs_Videau}. As a result, the incident angle of the particles is no longer normal to the detector layers in the second module. To account for this effect, the angular correction described in section~\ref{angular_correction_chapter} is applied to hits in the second module. In the following sections, unless stated otherwise, $N_i$ therefore refers to the angle-corrected number of hits.

\subsection{Energy reconstruction methods}
\label{Calibration_procedure}

This section describes the different ways of reconstructing the SDHCAL energy. The results obtained for each method on both samples of single $K^0_L$ and dijet events are presented in section~\ref{Results}. The number of significant digits reported for all calibration parameters throughout this section reflects the direct output of the minimization procedure. No rounding was applied in order to ensure full reproducibility of the results within the reconstruction framework. The associated parameter uncertainties are not reported, as the coefficients are intended for direct use in the reconstruction and their individual uncertainties are not relevant for this purpose, although they could be considered in future studies dedicated to systematic effects.

\subsubsection{Linear reconstruction}
\label{linear_reco_chapter}

As described in section~\ref{simulation_section}, the energy of the initial particle can be reconstructed as a function of $N_1$, $N_2$ and $N_3$. A first possibility is to define the reconstructed energy as a weighted sum of these numbers of hits:

\begin{equation}
\label{original_reco_formula}
    E_{\text{reco}} = \alpha N_1+\beta N_2 + \gamma N_3
\end{equation}

The calibration procedure then consists in finding the best way to compute $\alpha, \beta, 
 \text{ and } \gamma$ to obtain the best reconstructed energy in terms of both linearity and resolution. 

The first and simplest approach is the linear energy reconstruction formula. In this configuration, $\alpha, \beta, \text{ and } \gamma$ are defined as constants. Each of these constants can be interpreted as the corresponding energy for each SDHCAL threshold in GeV, and are thus called energy factors. The linear reconstruction formula is the default one used by PandoraPFA, as it has been mostly developed for analog calorimeters. 

The parameters for the linear energy reconstruction formula are adjusted by minimizing a $\chi^{2}$-like expression with \texttt{TMinuit} and \texttt{MIGRAD} from \texttt{ROOT}~\cite{BrunRademakers1997ROOT}, based on the \texttt{MINUIT} minimization package~\cite{JamesRoos1975Minuit}.

\begin{equation}
\label{chi2_equation}
    \chi^2 = \frac{1}{N} \sum_{i=1}^{N}\frac{\left( E_{\text{mc}}^i - E_{\text{reco}}^i \right)^2}{\sigma_{i}^2}
\end{equation}
where $N$ is the number of events used to perform the minimization and $\sigma_i =\sqrt{E_{\text{mc}}^i}$~\cite{FirstResultsSDHCAL}. 
The minimization is performed using the calibration dataset described in section~\ref{Calibration_dataset}, yielding a minimum value of $\chi^2\simeq1.43$. 

The values of the parameters for the linear energy reconstruction formula can be found in table~\ref{Tab_linear}. 

\begin{table}[h]
\centering
\caption{Values of the parameters for the linear energy reconstruction formula}
\begin{tabular}{ccc}
Parameter & Value \\ \hline
$\alpha$     & $3.67023\times10^{-2}$  \\
$\beta$      & $7.45279\times10^{-2}$  \\
$\gamma$     & $3.63042\times10^{-1}$  
\end{tabular}
\label{Tab_linear}
\end{table}

\subsubsection{Quadratic reconstruction}
\label{Quad_reco}

Since the density of particles in a shower tends to increase with energy, using the linear energy formula may not be sufficient to take into account the saturation effect that increases at higher jet energies. It has thus been proposed to still use Formula \ref{original_reco_formula} but to introduce non-linearity by parametrizing $\alpha, \beta, \text{ and } \gamma$ as quadratic functions of the total number of hits in the HCAL ($N_{\text{hit}}=N_1+N_2+N_3$)~\cite{FirstResultsSDHCAL}. Using this formula implies the need to adjust 9 parameters in total, as shown in Formula~\ref{quad_formula}.

\begin{equation}
\label{quad_formula}
    \begin{array}{rcl}
\alpha  = \alpha_1 + \alpha_2\,N_{\text{hit}} + \alpha_3\,N_{\text{hit}}^2 \\
 \beta  = \beta_1 + \beta_2\,N_{\text{hit}} + \beta_3\,N_{\text{hit}}^2 \\
 \gamma  = \gamma_1 + \gamma_2\,N_{\text{hit}} + \gamma_3\,N_{\text{hit}}^2
\end{array}
\end{equation}

These parameters are adjusted using the same $\chi^2$-like expression as the linear formula. The minimization achieves a minimum value of $\chi^2\simeq1.35$. 

\begin{table}[h]
\centering
\caption{Values of the parameters for the quadratic energy reconstruction formula.}
\begin{tabular}{cc}
Parameter & Value \\ \hline
$\alpha_1$ & $6.30\times10^{-2}$ \\
$\alpha_2$ & $-1.17\times10^{-4}$ \\
$\alpha_3$ & $7.1818\times10^{-8}$ \\ \hline
$\beta_1$  & $4.10\times10^{-2}$ \\
$\beta_2$  & $1.39\times10^{-4}$ \\
$\beta_3$  & $-9.17877\times10^{-8}$ \\ \hline
$\gamma_1$ & $1.0\times10^{-10}$ \\
$\gamma_2$ & $1.105\times10^{-3}$ \\
$\gamma_3$ & $-6.43333\times10^{-7}$
\end{tabular}
\label{Tab_quad}
\end{table}

The values of the parameters for the quadratic energy reconstruction formula are given in table~\ref{Tab_quad}. The fit is first performed without constraints on the parameters. If the value of $\gamma_1$ is negative, it may lead to unphysical negative reconstructed energies for some configurations. In this case, the fit is redone with an imposed lower bound of $1\times10^{-10}$~GeV to $\gamma_1$. For the results reported in table~\ref{Tab_quad}, this lower bound was applied, leading to the fitted value of $\gamma_1$ quoted in the table.

\subsubsection{Split method}
\label{split_method_section}

Looking at the results in figure~\ref{KLong_results} for the single $K^0_L$ samples, the two classical formulas described earlier achieve good average linearity, but the linear reconstruction tends to overestimate the true energy, while the quadratic reconstruction tends to underestimate it. Linearity also deteriorates below 20~GeV. To address these issues, two separate quadratic functions were introduced instead of a single one. 

\begin{table}[h]
\centering
\caption{Values of the parameters for the split energy reconstruction formula.}
\begin{tabular}{ccc}
Parameter & Value ($N_{\text{hit}} \leq 400$) & Value ($N_{\text{hit}} > 400$)  \\ \hline
$\alpha_1$ & $7.12495\times10^{-2}$ & $3.50654\times10^{-2}$ \\
$\alpha_2$ & $-1.86449\times10^{-4}$ & $-3.84159\times10^{-5}$ \\
$\alpha_3$ & $1.97156\times10^{-7}$ & $2.40987\times10^{-8}$ \\ \hline
$\beta_1$  & $1.53121\times10^{-2}$ & $1.37626\times10^{-1}$ \\
$\beta_2$  & $4.5658\times10^{-4}$ & $-1.2783\times10^{-4}$ \\
$\beta_3$  & $-6.3797\times10^{-7}$ & $7.27967\times10^{-8}$ \\ \hline
$\gamma_1$ & $1.28742\times10^{-12}$ & $1.0\times10^{-10}$ \\
$\gamma_2$ & $1.00913\times10^{-3}$ & $1.13873\times10^{-3}$ \\
$\gamma_3$ & $-5.59527\times10^{-7}$ & $-6.63919\times10^{-7}$
\end{tabular}
\label{Tab_split}
\end{table}

In this approach, one quadratic function is fitted using events with a low number of hits, corresponding predominantly to low-energy showers, and is subsequently applied to small clusters. A second function is similarly fitted using events with a high number of hits and applied to large clusters. As shown in figure~\ref{N1N2N3vsEnergy}, a threshold of 20~GeV roughly corresponds to 400 hits in the SDHCAL. Various transition thresholds, ranging from 200 to 600 hits, were tested. 
Ultimately, the optimal results were achieved by applying the function fitted on events with $N_{\text{hit}}\leq600$ to clusters with $N_{\text{hit}}\leq400$, and the function fitted on events with $N_{\text{hit}}>400$ to clusters with $N_{\text{hit}}>400$. Here, the calibration is performed using the total number of hits at the event level, while the reconstructed energy is evaluated at the cluster level within the PFA reconstruction. For single $K^{0}_{L}$ events, both quantities are expected to coincide in most cases, although small differences may arise due to clustering effects such as fragmentation. The overlap between the fitting ranges ensures a smooth transition from the small cluster function to the large cluster function.
The values of the parameters for the split method can be found in table~\ref{Tab_split}.
The same fitting procedure as in section~\ref{Quad_reco} is applied to the two parametrizations. The corresponding minimizations yield $\chi^2\simeq1.60$ for the low-$N_{\text{hit}}$ region and $\chi^2\simeq0.62$ for the high-$N_{\text{hit}}$ region. Because of the $N_{\text{hit}}$ selection, these values are obtained on different event subsets and are not directly comparable to the global minimization results of the other methods.

\subsubsection{Polynomial regression}

As machine learning algorithms are becoming increasingly common, a method based on machine learning tools has been tested for hadronic energy reconstruction. This method uses \texttt{scikit-learn}, an open-source library for machine learning in Python~\cite{scikit-learn}. Although the procedure relies on machine learning tools, the resulting model is essentially a polynomial regression aimed at reconstructing the hadronic energy.

The procedure consists of training an algorithm on the dataset of single $K^0_L$ events previously used, split into training and testing samples with a 0.8/0.2 ratio. The input variables (features) are the numbers of hits in the SDHCAL, denoted $N_1$, $N_2$, and $N_3$, while the target variable (label) to be reconstructed is the particle true energy. As before, only events fully contained within the SDHCAL are retained for both training and testing.

The \texttt{PolynomialFeatures} tool is then applied to ensure that the output is a polynomial combination of all possible terms up to a chosen degree. The resulting formula can be written in its general form as:

\begin{equation}
E_{\text{reco}} =
\sum_{i=1}^{3} \alpha_i N_i
+ \sum_{i=1}^{3}\sum_{j=i}^{3} \beta_{ij} N_i N_j
+ \cdots
\end{equation}

This formula is then validated on the test sample. Tests conducted with different polynomial degrees showed that degree 2 provides the best compromise between reconstruction performance and model complexity. Consequently, the results reported here use a degree-2 polynomial, leading to nine coefficients to be determined.

\begin{table}[h]
\centering
\caption{Values of the parameters for the quadratic polynomial regression formula.}
\begin{tabular}{cc}
Parameter & Value \\ \hline
$\alpha_1$ & $3.0334 \times 10^{-2}$ \\ 
$\alpha_2$ & $8.6529 \times 10^{-2}$ \\
$\alpha_3$ & $3.20575 \times 10^{-1}$ \\ \hline 
$\beta_{11}$ & $1.4 \times 10^{-5}$ \\
$\beta_{12}$ & $1.90 \times 10^{-4}$ \\
$\beta_{13}$ & $-9.69 \times 10^{-4}$ \\
$\beta_{22}$ & $-4.03 \times 10^{-4}$ \\
$\beta_{23}$ & $5.00 \times 10^{-4}$ \\
$\beta_{33}$ & $4.948 \times 10^{-3}$
\end{tabular}
\label{Tab_poly_coeff}
\end{table}

The values of these coefficients are reported in table~\ref{Tab_poly_coeff}. An analogous split method to the one described in section~\ref{split_method_section} was tested using two polynomial combinations, but it did not yield any significant improvement over the single polynomial approach.

\section{Results}
\label{Results}

\subsection{Single $K^0_L$ reconstruction}

All the methods described in section~\ref{Calibration_procedure} were first tested on the same samples of single $K^0_L$ events used to fit the formulas. However, for the performance evaluation presented in this section, the containment and pure-HCAL selection criteria applied to define the calibration dataset are not imposed. The studied samples therefore also include events with energy deposited in the ECAL as well as events affected by leakage. In addition, while the calibration procedure is performed directly at the event level using the number of SDHCAL hits without PFA reconstruction, the results presented here are obtained after full PFA reconstruction with APRIL. For these samples, the analysis is based on the following procedure. 

For each energy point, the histogram of the total energy of the Particle Flow Objects (PFOs) is first fitted with a  Gaussian function on the whole distribution, giving $\mu_1$ and $\sigma_1$. A second Gaussian fit is then performed in the interval $\left[ \mu_1-1.5\,\sigma_{1}, \mu_1+1.5\,\sigma_{1} \right]$, giving $\mu_2$ and $\sigma_2$. These parameters are taken as the mean reconstructed energy, $E_{\text{reco}}$, and the event by event uncertainty on the reconstructed energy, $\sigma_{E_{\text{reco}}}$. 
The resolution is then defined as $R=\frac{\sigma_{E_{\text{reco}}}}{E_{\text{reco}}}$ and the linearity as: 
\begin{equation}
    \Delta E/E_{\text{mc}}=\left( E_{\text{reco}} - E_{\text{mc}} \right)/E_{\text{mc}}
\end{equation}

\begin{figure}[htbp]
  \centering
  % Figure a
  \subfloat[Linearity \label{LinearityKlong}]{
    \includegraphics[width=0.48\textwidth]{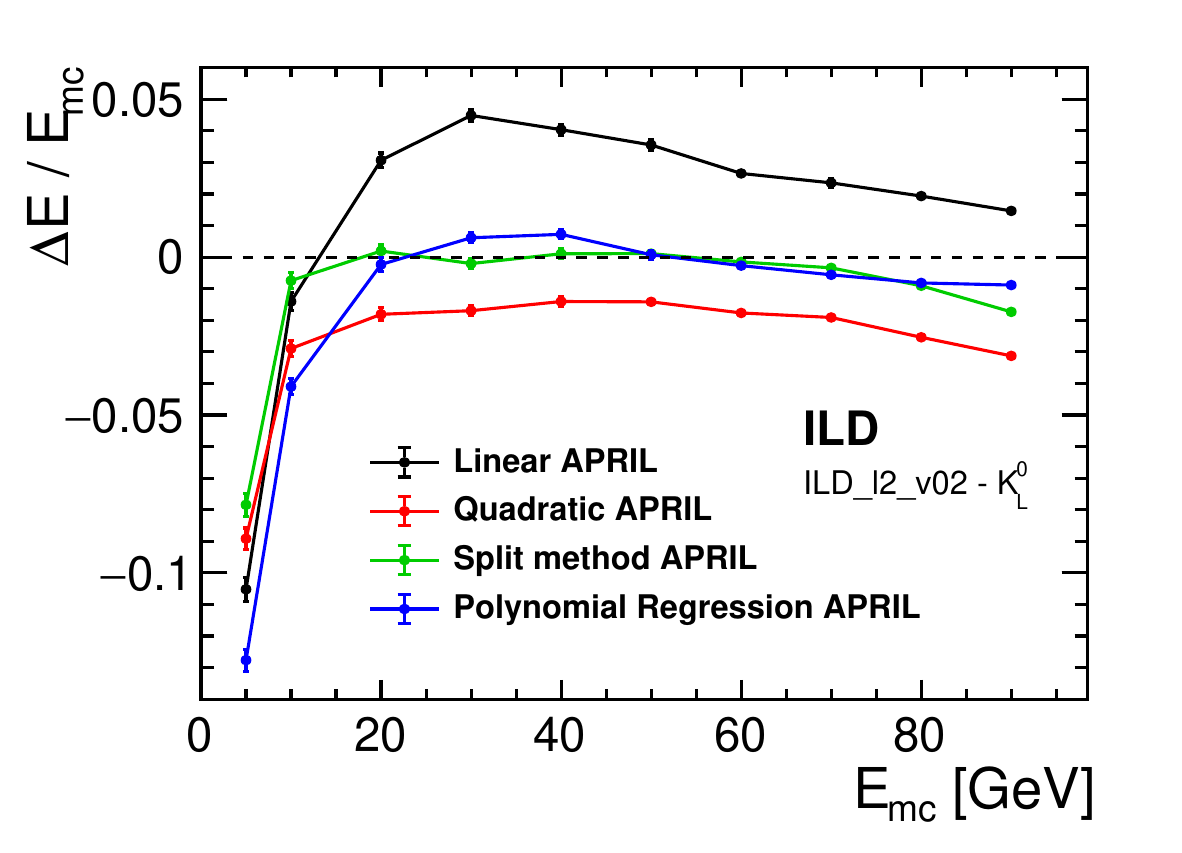}
  }
  \hfill
  % Figure b
  \subfloat[Resolution \label{ResolutionKlong}]{
    \includegraphics[width=0.48\textwidth]{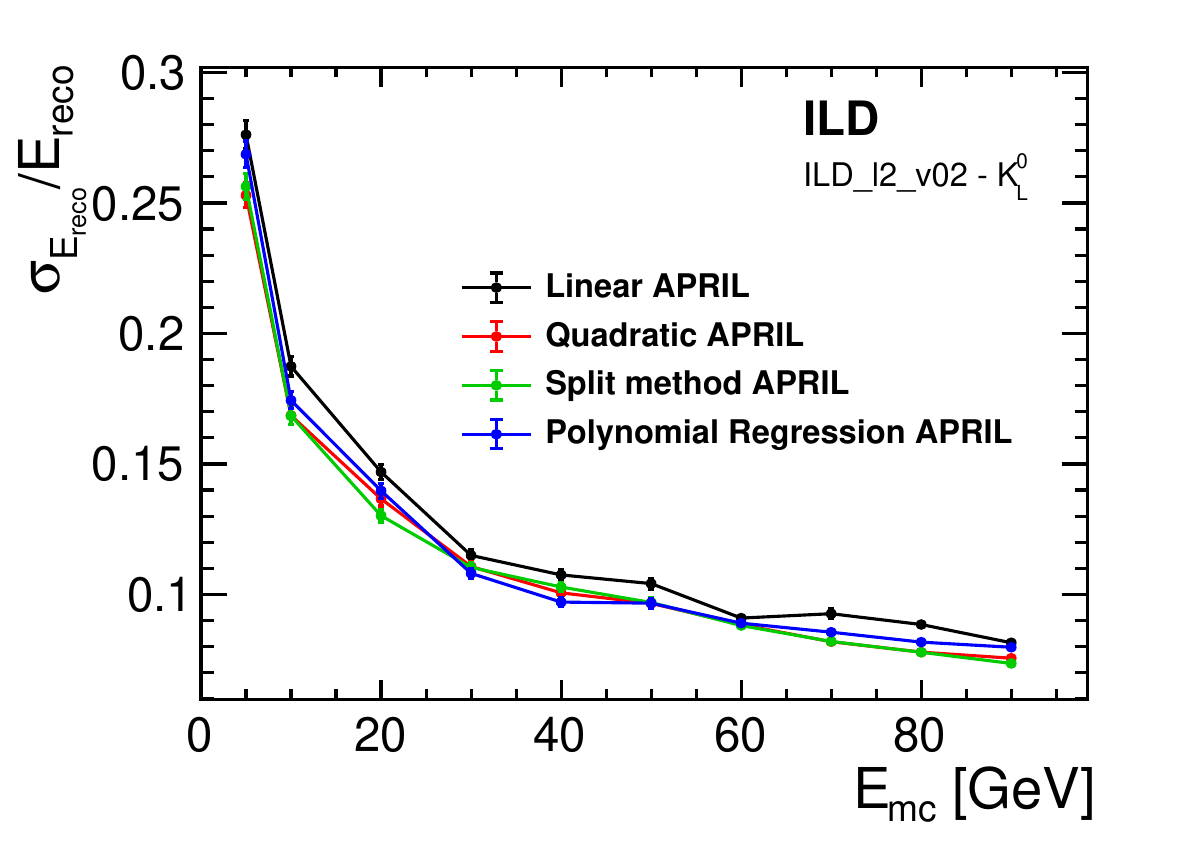}
  }
  \caption{Linearity (left) and resolution (right) for the different reconstruction formulas on samples of single $K^0_L$ sent into the barrel with no angle and reconstructed with APRIL.}
  \label{KLong_results}
\end{figure}

\begin{figure}[h]
    \centering
    \includegraphics[width=0.52\linewidth]{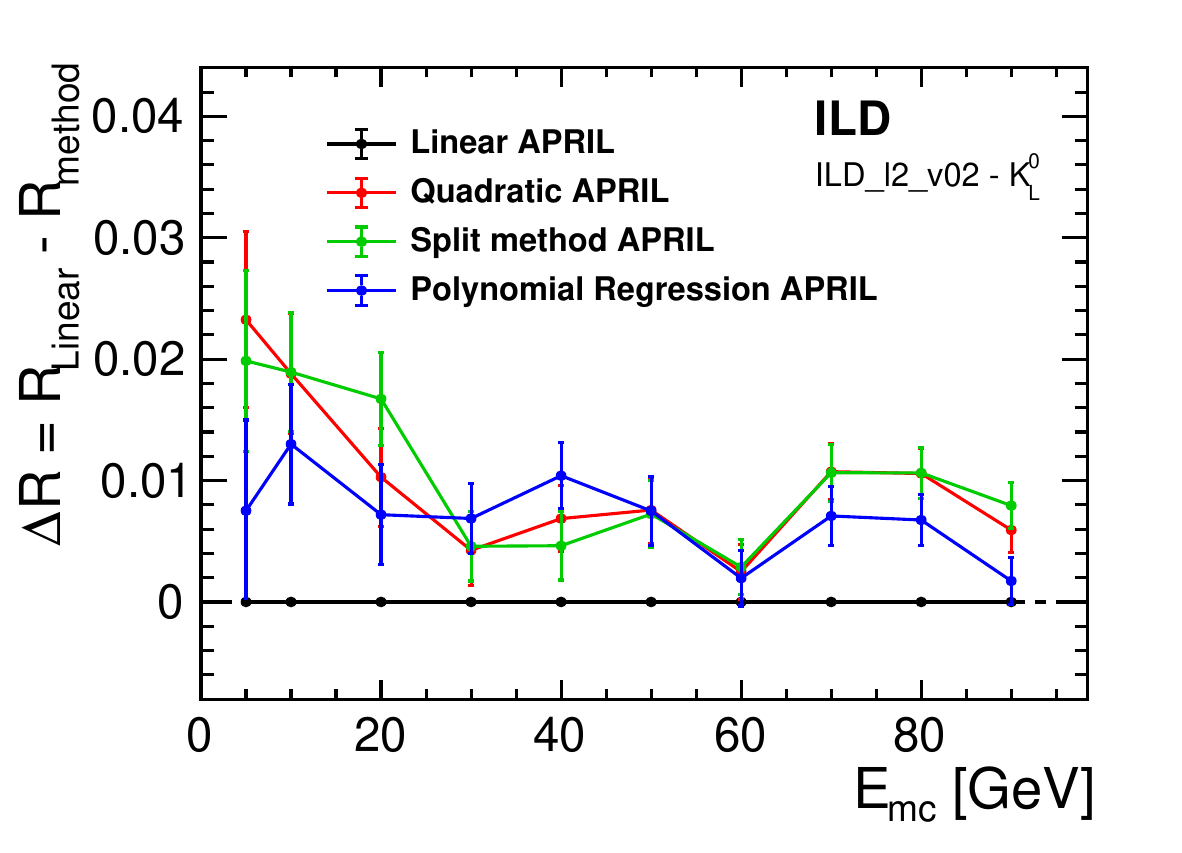}
    \caption{Difference between the resolution obtained with the linear method and that obtained with other reconstruction methods for single $K^{0}_L$ samples reconstructed with APRIL.}
    \label{diff_resolution_klong}
\end{figure}

The results obtained are shown in figure~\ref{KLong_results} and figure~\ref{diff_resolution_klong}. 
Above 10~GeV, the linear method exhibits the largest deviation from linearity among the studied approaches, with deviations ranging from approximately +1\% to +5\%. The non-linear methods show smaller deviations over the same energy range, typically between 0\% and -3.5\%. Among them, the split method provides the smallest deviations, remaining within 0\% to -2\%.
In terms of energy resolution, the split method yields improved performance between 5~GeV and 20~GeV compared to the other approaches. The linear method systematically results in the poorest resolution across all investigated energy points.
This behavior is consistent with expectations for such a simple parametrization, as it is harder to capture the complexity and evolution of the hadronic shower structure over the full energy range with $\alpha, \beta, \text{ and } \gamma$ defined as constants. 
The observed performance is consistent with results obtained from SDHCAL test beam data using charged pion beams~\cite{FirstResultsSDHCAL}. Despite the different context, similar behaviors are observed. In particular, a linearity better than 5\% is achieved above 10~GeV in both cases. The energy resolution is also found to be comparable, with values around 25\% at 5~GeV and about 7--8\% at 80~GeV.

\subsection{Dijet reconstruction}

After testing the different formulas on the single $K^0_L$ samples, they were applied to more complex physics cases. As it is common to use such events to benchmark the performance of PFAs, samples of dijet ($u, d, s$) with energies ranging from 30~GeV to 280~GeV were used. Unlike the samples of single $K^0_L$, these dijet events are simulated with a uniform angular distribution, but a cut on the angle of $|\cos \theta| < 0.7$ is applied in the analysis to ensure that the jets are contained in the barrel. This prevents the results from being too affected by the non-instrumented areas due to the gap between the barrel and the endcaps. Unless stated otherwise, results presented here are obtained after full PFA reconstruction with APRIL.

\begin{figure}[htbp]
  \centering
  % Figure a
  \subfloat[Linearity\label{Linearity_uds}]{
    \includegraphics[width=0.48\textwidth]{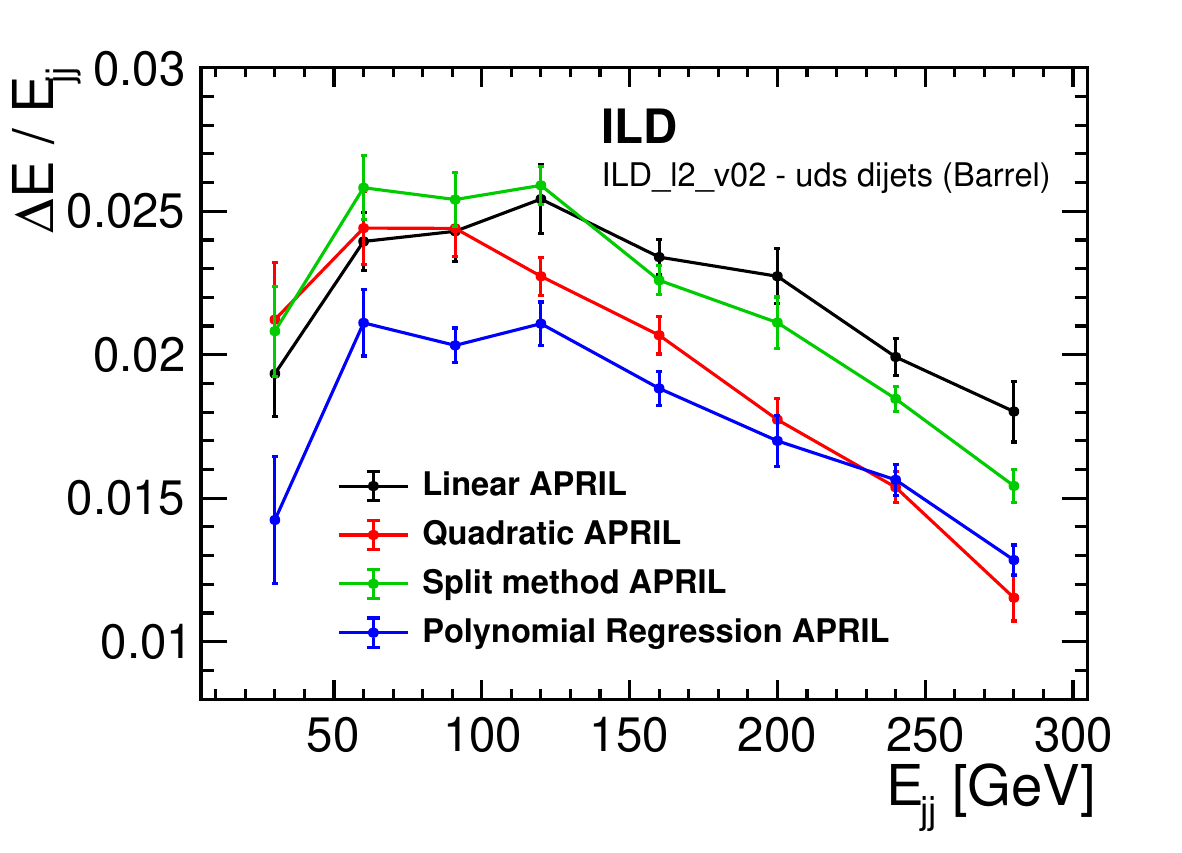}
  }
  \hfill
  % Figure b
  \subfloat[Resolution\label{Resolution_uds}]{
    \includegraphics[width=0.48\textwidth]{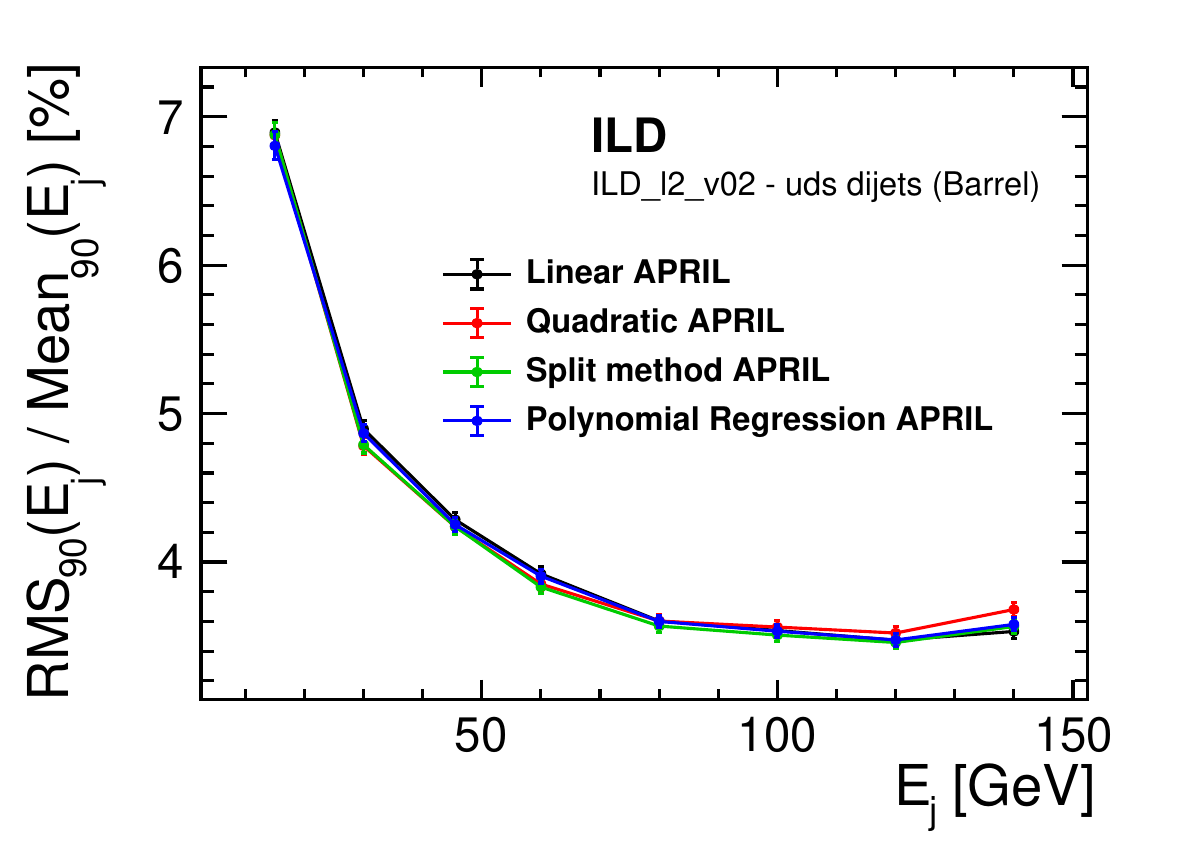}
  }
  \caption{Linearity (left) and resolution (right) for the different reconstruction formulas on samples of dijet events ($u,d,s$) generated at various angles with $|\cos \theta| < 0.7$ and reconstructed with APRIL.}
  \label{uds_results}
\end{figure}

Each point in figure~\ref{uds_results} is obtained by computing the $\mathrm{Mean}_{90}(\mathrm{E}_{\mathrm{jj}})$ and $\mathrm{RMS}_{90}(\mathrm{E}_{\mathrm{jj}})$ of the total PFO energy distribution, with $\mathrm{E}_{\mathrm{jj}}$ being the energy of the jet pair. Here, $\mathrm{Mean}_{90}(\mathrm{E}_{\mathrm{jj}})$ and $\mathrm{RMS}_{90}(\mathrm{E}_{\mathrm{jj}})$ are respectively the mean value and the root mean square computed over the smallest interval containing at least 90\% of the events, in order to reduce the impact of non-Gaussian tails. If $\mathrm{E}_{\mathrm{j}}$ is the energy of a single jet, the jet energy resolution (JER) can then be computed as: 
\begin{equation}
    \text{JER}=\frac{\text{RMS}_{90}(\text{E}_\mathrm{j})}{\text{Mean}_{90}(\text{E}_\mathrm{j})}
= \sqrt{2} \cdot\frac{\mathrm{RMS}_{90}(\mathrm{E}_{\mathrm{jj}})}{\mathrm{Mean}_{90}(\mathrm{E}_{\mathrm{jj}})}
\end{equation}

The linearity is here defined as: 
\begin{equation}
    \Delta E/\mathrm{E}_{\mathrm{jj}}=\left( \mathrm{Mean}_{90}(\mathrm{E}_{\mathrm{jj}}) - \mathrm{E}_{\mathrm{jj}} \right)/\mathrm{E}_{\mathrm{jj}}
\end{equation}
The linearity achieved by all reconstruction methods is in the range [+1.1\%, +2.6\%] over the full energy spectrum. Among them, the polynomial regression provides the best overall linearity, as shown in figure~\ref{uds_results}\subref{Linearity_uds}. 
Regarding the jet energy resolution, the performances of the different methods are overall very similar, as illustrated in figure~\ref{uds_results}\subref{Resolution_uds}.

For comparison, the jet energy resolution achieved in the standard ILD setup, using the scintillator based readout option for the HCAL (AHCAL) in the TESLA geometry, and reconstructed with PandoraPFA, reaches values below 4\% in the barrel at the $Z$ pole ($\sqrt{s}=91$~GeV), and improves to below 3\% for jet energies above 100~GeV~\cite{ILDConceptGroup:2020sfq}. The resolutions obtained in this study with the SDHCAL and APRIL are therefore slightly worse than the ILD benchmark performance. The observed differences with respect to this benchmark are primarily attributed to the reconstruction strategy rather than to the calorimeter technology alone. In particular, PandoraPFA includes a highly optimized reclustering procedure that significantly reduces confusion effects at high jet energies. The current APRIL reconstruction does not yet include equivalent reclustering capabilities, which can explain part of the observed degradation in this regime. Differences in detector geometry may also contribute at a secondary level. In particular, the presence of non-instrumented gaps between the barrel wheels of the HCAL in the Videau configuration used in this study, whose impact can be seen in figure~\ref{fig_correction_hits}\subref{SubfigCosTheta}, is expected to degrade the performance. Such gaps are not present in the standard ILD geometry based on the TESLA design. The linearity, however, is found to be comparable to that achieved in the standard ILD configuration over the same energy range.

\begin{figure}[h]
    \centering
    \includegraphics[width=0.5\linewidth]{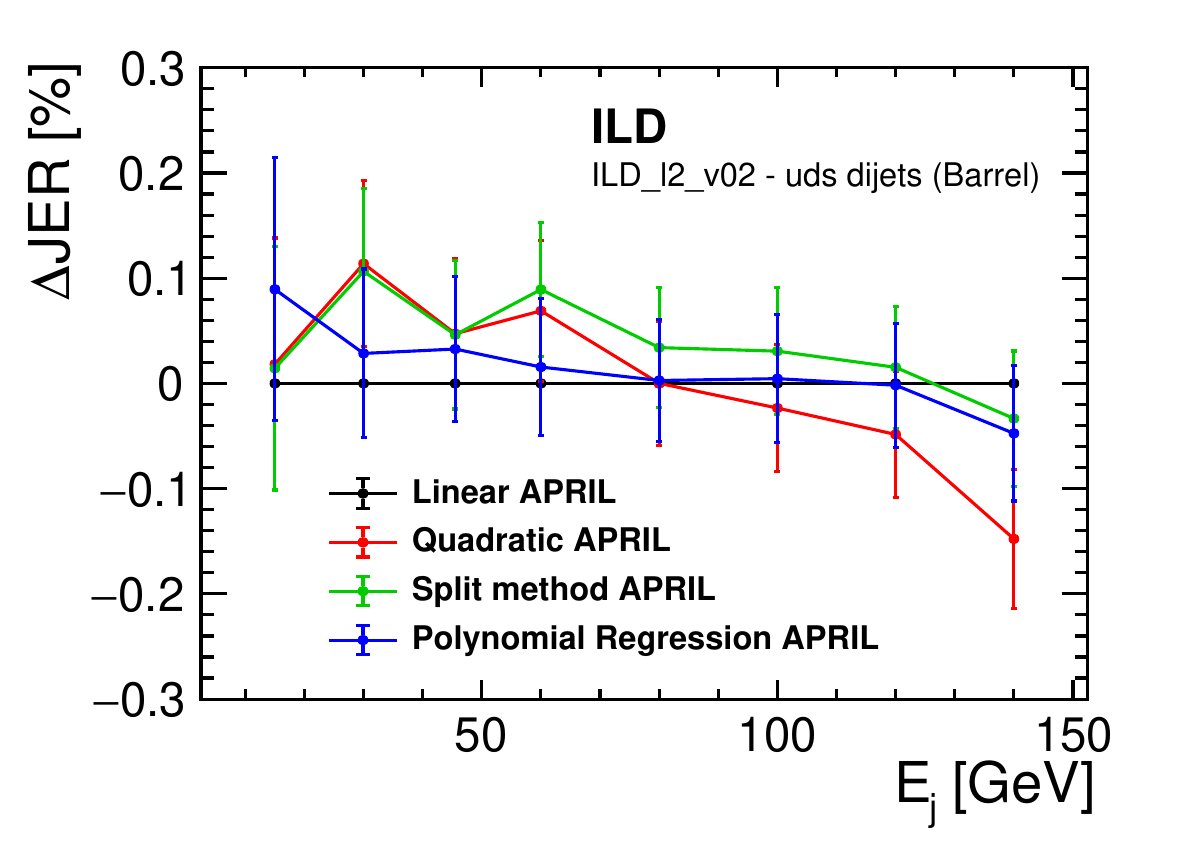}
    \caption{Difference between the jet energy resolution obtained with the linear method and that obtained with other methods, for dijet ($u$, $d$, $s$) 
samples with $|\cos \theta| < 0.7$ reconstructed with APRIL.}
    \label{diff_resolution_uds}
\end{figure}

To better resolve the variations between the different methods in the barrel, the linear reconstruction is taken as a reference and $\Delta\mathrm{JER} = \mathrm{JER}_{\mathrm{linear}} - \mathrm{JER}_{\mathrm{method}}$ is shown in figure~\ref{diff_resolution_uds}, allowing differences between methods of the order of a tenth of percent to be distinguished. Some notable variations with the jet energy can be observed for the quadratic reconstruction formula with respect to the linear one.

In fact, for $\mathrm{E}_\mathrm{{j}}<60$~GeV, figure~\ref{diff_resolution_uds} shows that the classical quadratic formula and the split method achieve a better resolution than the linear formula and the polynomial regression. However, at higher energies ($\mathrm{E}_\mathrm{{j}}\geq90$~GeV), this trend reverses: the quadratic formula gives the worst resolution, whereas the linear method provides one of the best.

Further investigations suggest that this effect arises from the PFA reconstruction itself rather than from the formula. Indeed, extensive studies performed for the ILD detector showed that confusion increases with increasing energy, and that for $\mathrm{E}_\mathrm{j} \geq 100$~GeV, the dominant contribution to the resolution is the confusion term~\cite{Green:confusion}. This confusion arises when a hit is not attributed to the correct initial particle by the reconstruction algorithm. Since the quadratic formula is more precise than the linear one for single clusters, it is more sensitive to fluctuations in the number of hits caused by confusion. Consequently, these errors accumulate over multiple clusters, leading to worse overall resolution. 

\begin{figure}[htbp]
  \centering
  % Figure a
  \subfloat[Linearity\label{linearity_perfect}]{
    \includegraphics[width=0.48\textwidth]{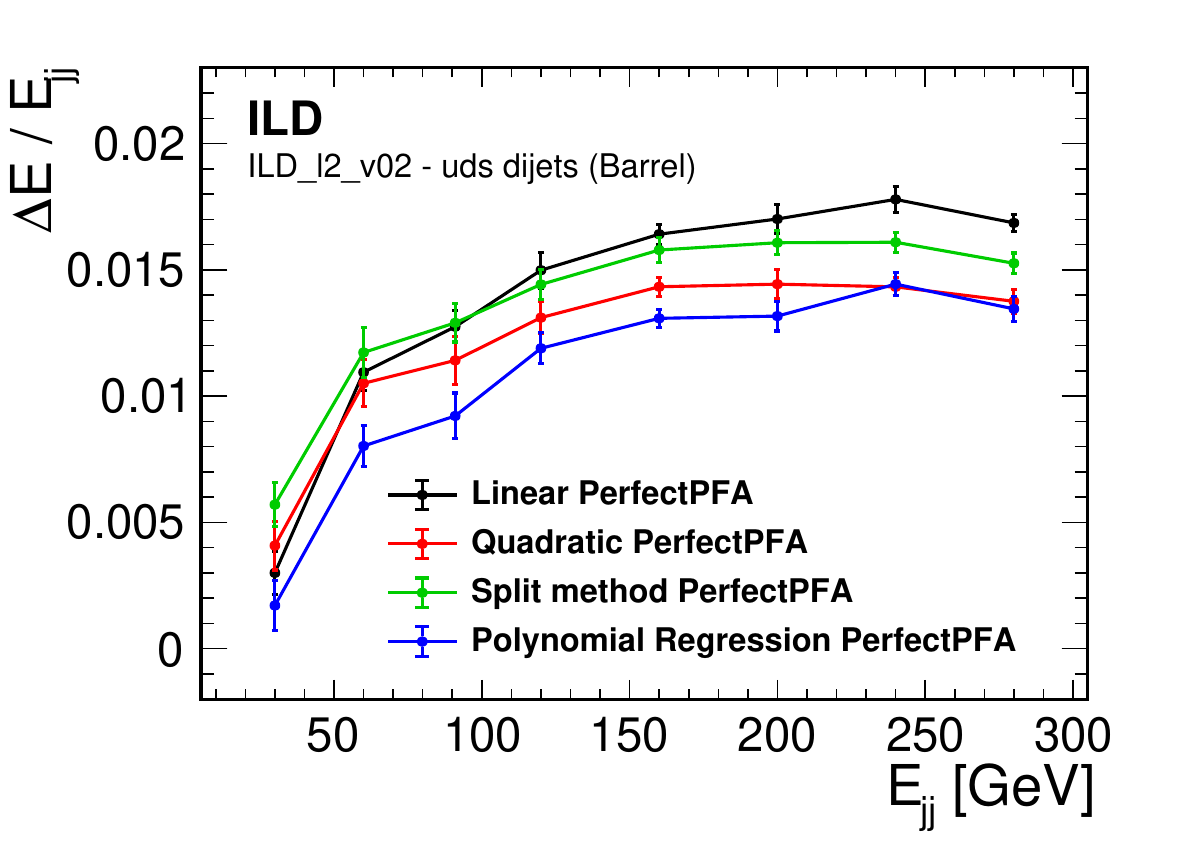}
  }
  \hfill
  % Figure b
  \subfloat[Resolution\label{resolution_perfect}]{
    \includegraphics[width=0.46\textwidth]{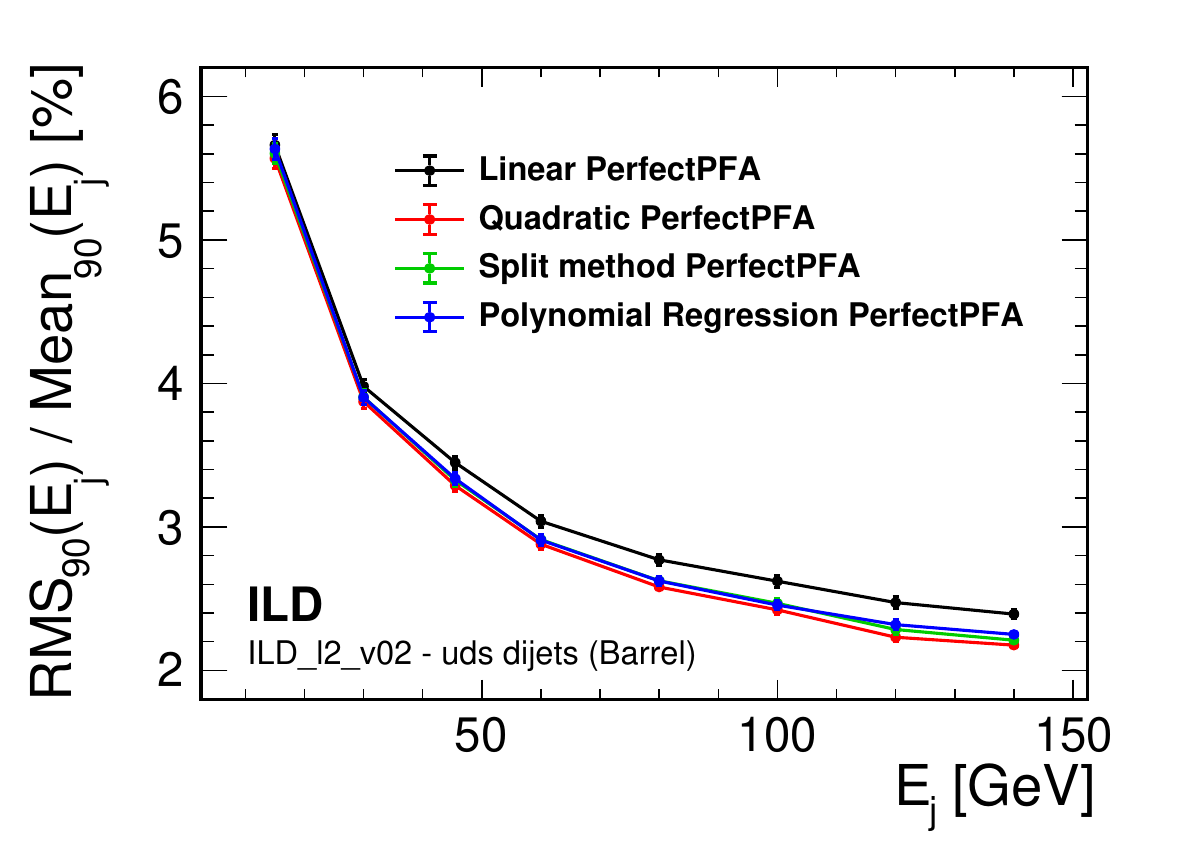}
  }
  \caption{Linearity (left) and resolution (right) for the different reconstruction formulas on samples of dijet events ($u,d,s$) generated at various angles with $|\cos \theta| < 0.7$ and reconstructed with PerfectPFA.}
  \label{PerfectPFA_resolution}
\end{figure}

\begin{figure}[h]
    \centering
    \includegraphics[width=0.5\linewidth]{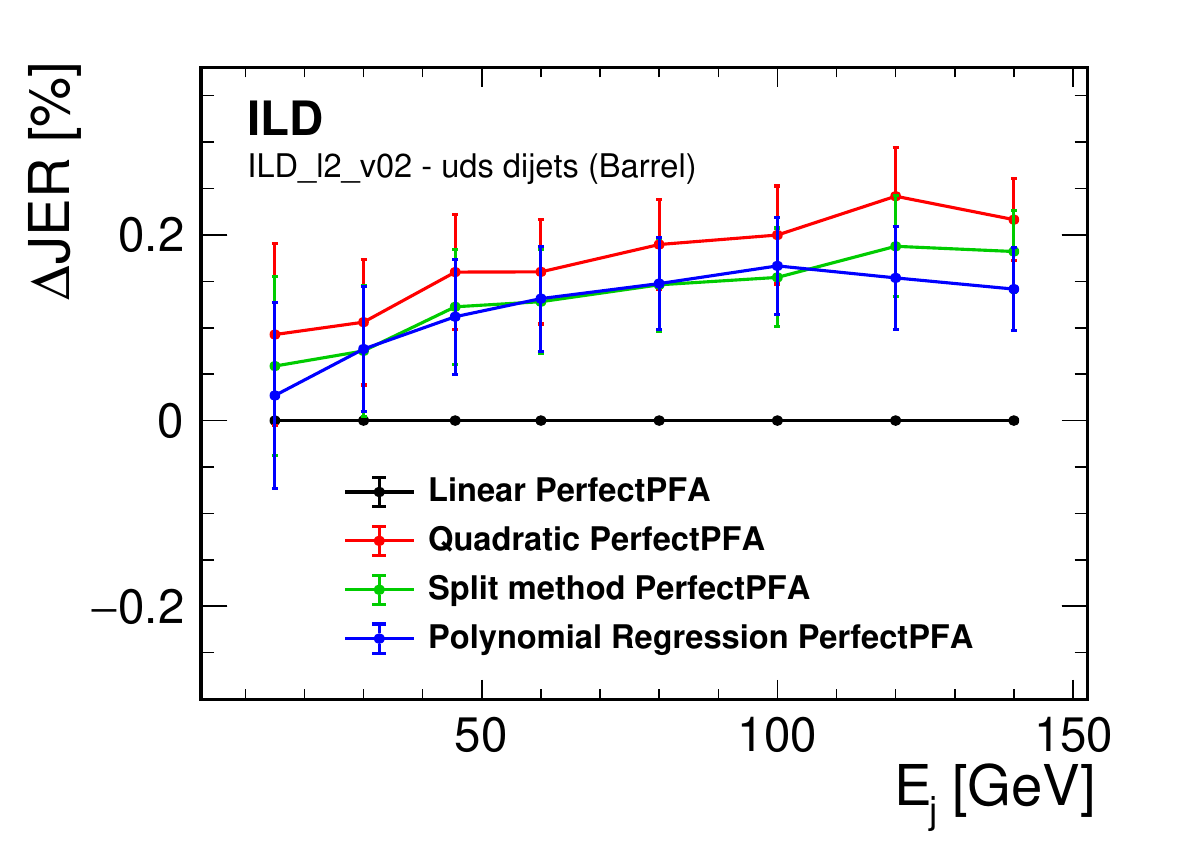}
    \caption{Difference between the jet energy resolution obtained with the linear method and that obtained with other methods, for dijet ($u$, $d$, $s$) samples reconstructed with PerfectPFA.}
    \label{diff_resolution_uds_perfect}
\end{figure}

This hypothesis can be confirmed by applying a PerfectPFA, which deliberately uses Monte-Carlo truth information to reconstruct particles. By doing so, the confusion term disappears as the particles are perfectly reconstructed and only the contribution from the detector's intrinsic resolution remains. As shown in figure~\ref{PerfectPFA_resolution}\subref{resolution_perfect} and figure~\ref{diff_resolution_uds_perfect}, when applied with PerfectPFA, all formulas perform better than the linear formula on the whole energy range. The linearity shown in figure~\ref{PerfectPFA_resolution}\subref{linearity_perfect} is within the range [+0.1\%, +1.8\%] and the polynomial regression still provides the best overall linearity. A similar confirmation is obtained when repeating the study with the standard PandoraPFA reconstruction, which includes dedicated reclustering procedures that mitigate confusion effects at high energies. In this case, the quadratic reconstruction remains superior to the linear one over the full energy range.

These findings show that the optimal choice depends on the physics case and the reconstruction quality: under realistic PFA-induced confusion at high jet energies, the linear formula becomes competitive, whereas with PerfectPFA (Monte-Carlo truth) or a nearly ideal PFA, non-linear methods provide superior performance across the full energy range. At low jet energies where confusion is less important, quadratic and split formulas give best results.

\begin{figure}[h]
  \centering
  % Figure a
  \subfloat[Linearity\label{Linearity_uds_endcap}]{
    \includegraphics[width=0.48\textwidth]{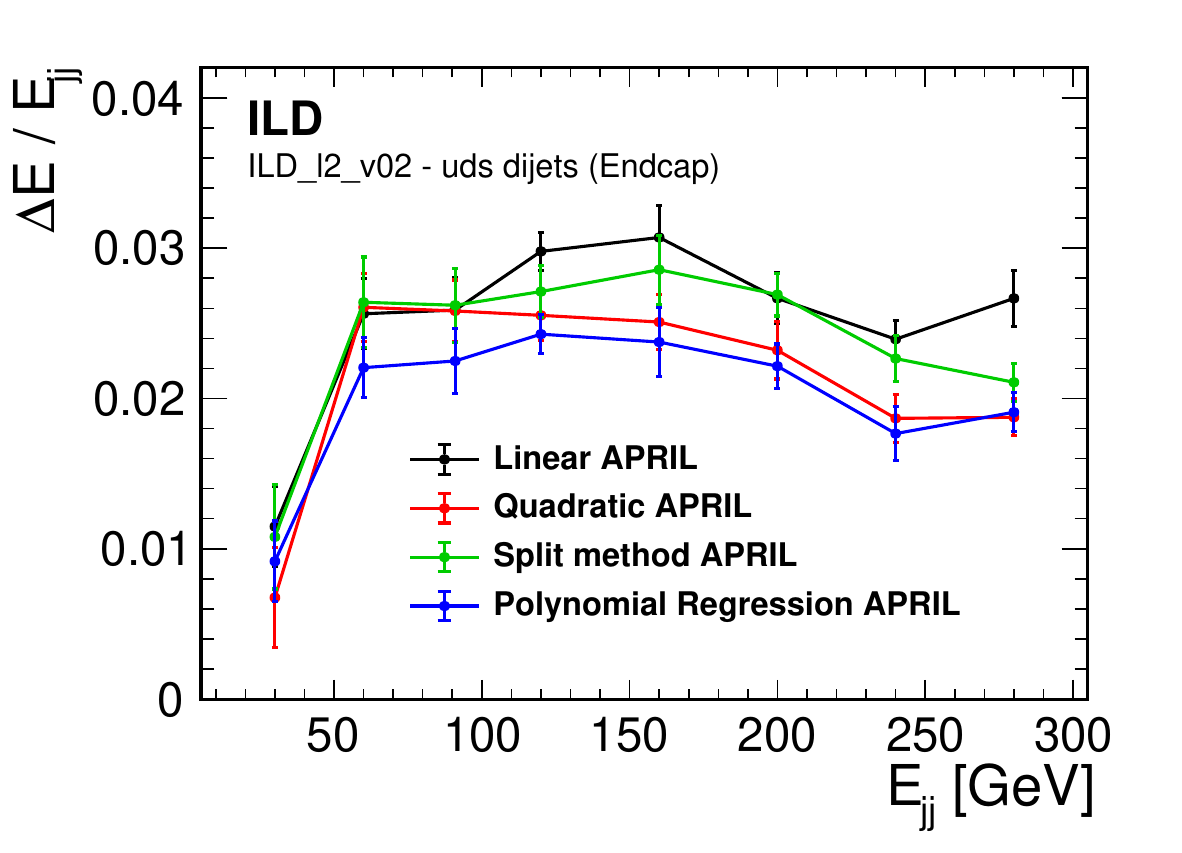}
  }
  \hfill
  % Figure b
  \subfloat[Resolution\label{Resolution_uds_endcap}]{
    \includegraphics[width=0.48\textwidth]{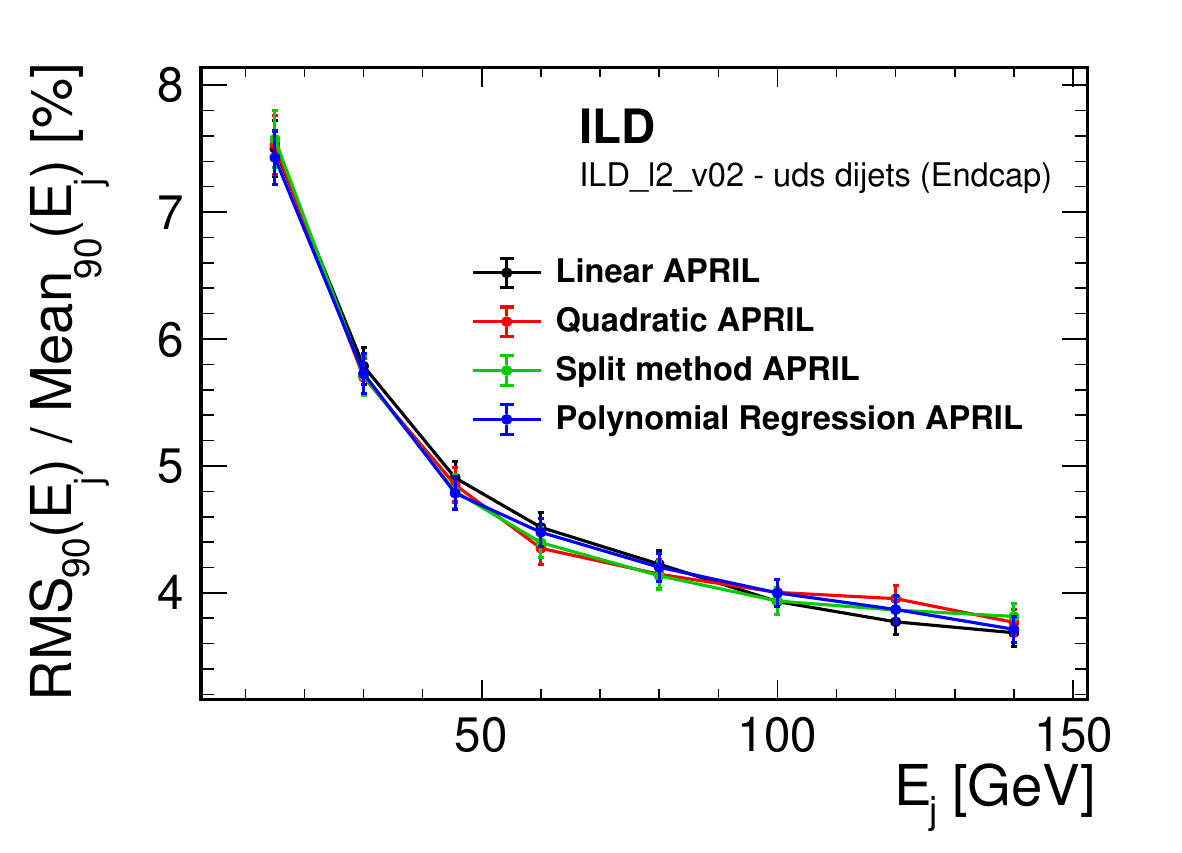}
  }
  \caption{Linearity (left) and resolution (right) for the different reconstruction formulas on samples of dijet events ($u,d,s$) generated at various angles with $0.8\le|\cos \theta| \le 0.9$ (endcap) and reconstructed with APRIL.}
  \label{uds_results_endcap}
\end{figure}

Although the analysis presented above focuses on jets contained in the barrel region, a similar study was performed for jets reconstructed in the endcap, with $0.8\le|\cos \theta| \le 0.9$. The results shown in figure~\ref{uds_results_endcap} display a consistent behavior, with linearity and resolution comparable to those observed in the barrel, and only a slight degradation in resolution, in line with what is also observed in the standard ILD benchmark~\cite{ILDConceptGroup:2020sfq}. This confirms that the reconstruction methods remain valid across different regions of the detector.

%% file: section3.tex
\section{Conclusion}

Several reconstruction methods for hadronic energy in the SDHCAL have been studied, ranging from simple linear parametrizations to quadratic and split formulas, as well as a machine-learning inspired approach. Their performances were assessed on both single $K^{0}_{L}$ samples and dijet events, providing complementary benchmarks of calorimeter and PFA performance.

A dedicated angular correction was shown to be essential to restore energy linearity in the barrel region, compensating for geometry-induced variations in sampling.

The results confirm that all methods achieve linearity at the level of a few percent over a wide energy range. For single neutral hadrons, a linearity better than 5\% is obtained above 10~GeV, with energy resolutions ranging from approximately 25\% at 5~GeV down to 7--8\% at 80~GeV, and the split method shows the most robust behaviour, improving resolution at low energies while preserving linearity at higher ones.

For jets, the interplay with PFA reconstruction becomes critical: at high energies, confusion dominates the resolution and favors the linear parametrization due to its reduced sensitivity to fluctuations, while with PerfectPFA the non-linear methods recover their expected superiority. These observations highlight the importance of accounting for both detector effects and reconstruction algorithms when choosing a calibration strategy. The jet energy resolution obtained in this study reaches values between 3.4--4\% above 100~GeV, demonstrating a competitive performance. This remains slightly above the ILD benchmark based on the scintillator HCAL and PandoraPFA, which achieves resolutions of about 3\% at similar energies, while a comparable level of linearity is maintained.

Overall, the split method and the polynomial regression offer the best compromise across the different scenarios studied, combining stability with good resolution. At the same time, the sensitivity of the results to PFA confusion suggests that future progress will come from improved reconstruction techniques. Several directions are being pursued in this regard. Incorporating precise timing provided by the T-SDHCAL into APRIL could help disentangle overlapping showers and reduce confusion at high energies. The introduction of reclustering, or equivalent cluster-splitting techniques within APRIL is also expected to further mitigate confusion effects. Finally, dedicated corrections of detector non-uniformities, such as the gaps between HCAL barrel wheels inherent to the Videau geometry used in this study, could bring additional improvements. Together, these developments could ultimately lead to more accurate hadronic energy reconstruction at future lepton colliders.